\documentclass[twocolumn,prb,superscriptaddress,showpacs]{revtex4}
\usepackage{graphicx}          
\usepackage{dcolumn}           
\usepackage{bm}                
\usepackage{color}
\usepackage[normalem]{ulem}

\begin{document}

\title{Many-body effects on the Rashba-type spin splitting in bulk bismuth tellurohalides}

\author{I.~P.~Rusinov}
\affiliation{Tomsk State University, 634050, Tomsk, Russia}

\author{I.~A.~Nechaev}
\affiliation{Donostia International Physics Center (DIPC), 20018 San Sebasti\'an/Donostia, Basque Country, Spain}
\affiliation{Tomsk State University, 634050, Tomsk, Russia}

\author{S.~V.~Eremeev}
\affiliation{Institute of Strength Physics and Materials Science, 634021, Tomsk, Russia}
\affiliation{Tomsk State University, 634050, Tomsk, Russia}

\author{C.~Friedrich}
\affiliation{Peter Gr\"{u}nberg Institut and Institute for Advanced Simulation, Forschungszentrum J\"{u}lich and JARA, D-52425 J\"{u}lich, Germany}

\author{S. Bl\"{u}gel}
\affiliation{Peter Gr\"{u}nberg Institut and Institute for Advanced Simulation, Forschungszentrum J\"{u}lich and JARA, D-52425 J\"{u}lich, Germany}

\author{E. V. Chulkov}
\affiliation{Donostia International Physics Center (DIPC), 20018 San Sebasti\'an/Donostia, Basque Country, Spain}
\affiliation{Departamento de F\'isica de Materiales and Centro Mixto
CSIC-UPV/EHU, Facultad de Ciencias Qu\'{\i}micas, Universidad del
Pais Vasco/Euskal Herriko Unibertsitatea, Apdo. 1072, 20080 San
Sebasti\'an/Donostia, Basque Country, Spain}

\date{\today}

\begin{abstract}
We report on many-body corrections to one-electron energy spectra of bulk bismuth tellurohalides---materials that exhibit a giant Rashba-type spin splitting of the band-gap edge states. We show that the corrections obtained in the one-shot $GW$ approximation noticeably modify the spin-orbit-induced spin splitting evaluated within density functional theory. We demonstrate that taking into account many-body effects is crucial to interpret the available experimental data.
\end{abstract}


\maketitle

\section{Introduction}

Recently, bismuth tellurohalides have been discovered as a new very promising class of materials for semiconductor spintronics applications.\cite{NatMat_BiTeI, Bahramy, BTX_PRL, Crepaldi_PRL_2012, Landolt, BiTeI_JETPLett} Due to inversion asymmetry of bulk crystal potential, these semiconductors show a giant Rashba-type spin splitting of the bulk states. A study of bismuth tellurohalides has been triggered by Ref.~\onlinecite{NatMat_BiTeI}, where the band structure of BiTeI has been investigated with angle-resolved photoemission spectroscopy (ARPES). The spin-split state experimentally observed in Ref.~\onlinecite{NatMat_BiTeI} was attributed to the bulk conduction band minimum (CBM) shifted downward due to band bending in the near-surface region and confined in the respective accumulation layer. The splitting of such a CBM was characterized by the momentum offset $k_R\sim0.052$~\AA$^{-1}$ and the Rashba energy $E_R\sim0.1$~eV. These experimental values for the Rashba parameters were found to be in close agreement with those obtained from \textit{ab initio} density functional theory (DFT) calculations of the bulk BiTeI band structure (0.05 \AA$^{-1}$ and 0.113 eV, respectively, as evaluated in Ref.~\onlinecite{NatMat_BiTeI}). This fact was interpreted in favor of a bulk nature of the observed state.

In a DFT study \onlinecite{BTX_PRL} of bulk BiTeCl, BiTeBr, and BiTeI, it was revealed that these polar layered semiconductors all exhibit similarly large Rashba-type spin splittings. Additionally, in Ref. \onlinecite{BTX_PRL} and later in Refs. \onlinecite{BiTeI_JETPLett} and \onlinecite{Eremeev_NJP_2013}, it was shown that a free-electron-like spin-orbit-split surface state emerges in the bulk band gap at the (0001) surface of all mentioned semiconductors (hereafter we mean the Te-terminated surface). Similarly to the bulk CBM, this surface state possesses a Rashba-type spin splitting and the corresponding spin texture. Quantitatively, e.g., in the case of BiTeI, the surface-state spin splitting is described theoretically by the Rashba parameters\cite{BiTeI_JETPLett} $E_R=0.121$~eV and $k_R=0.068$~\AA$^{-1}$. The existence of this surface state has been experimentally corroborated in
Refs.~\onlinecite{Crepaldi_PRL_2012} and \onlinecite{Landolt}. Additionally, in these works it was questioned if the state observed by the ARPES in Ref.~\onlinecite{NatMat_BiTeI} was really a bulk-derived state confined in the band-bending accumulation layer (see also Ref.~\onlinecite{Martin_Arxiv_2012}). However, when comparing the original ARPES measurements with the above theoretical Rashba parameters for the surface state we find that they disagree. As to BiTeBr and BiTeCl, their surface electronic structure has been investigated by ARPES very recently in  Ref.~\onlinecite{Sakano_Arxiv_2013}. Analogously to BiTeI, signatures of spin-split states of a Rashba-type nature have been observed and interpreted in a similar way as in Ref.~\onlinecite{NatMat_BiTeI}. For these states, unlike BiTeI, experimental values of the Rashba parameters reported in  Ref.~\onlinecite{Sakano_Arxiv_2013} cannot be supported well by the available theoretical data obtained for the bulk conduction-band bottom.

When comparing the experimental ARPES data to band structures calculated within a Kohn-Sham (KS) DFT scheme one should keep in mind that, strictly speaking, the KS energies cannot be interpreted as excitation energies. For example, the fact that the KS band gap underestimates the experimental one often by a factor of two is known as the \textit{band-gap problem} of DFT. Also, the dispersion of its edges can be affected. An effective single-particle (quasiparticle) description of the many-body system can be retained if one employs an electronic self-energy that embodies the many-body exchange and correlation effects. Already on the level of the $GW$ approximation for the self-energy, calculations of many-body corrections to KS bands allow one to significantly improve theoretical  results (see, e.g., Ref. \onlinecite{GW_semiconds}). However, \textit{ab initio} $GW$ calculations aimed at a study of the surface electronic structure of materials, where spin-orbit interaction plays an important role, are not feasible so far.

In this paper, we present a theoretical \textit{ab initio} study of the bulk band structure of BiTeX (X$=$I, Br, Cl) taking into account many-body
corrections evaluated within the $GW$ approximation. We also distinguish between the local-density (LDA) and the generalized gradient approximation
(GGA) for the exchange-correlation functional used in the calculation of the KS reference system. First, the quasiparticle corrections to the KS bulk states will provide an answer to the question of whether the corrected values of the bulk CBM Rashba parameters still match the values coming from the experimental observation. Second, the fact that the surface-state properties are predetermined by the properties of the bulk band-gap edges\cite{BTX_PRL} allows one to estimate the effect of many-body corrections on the surface-state dispersion on the basis of the $GW$ results for the bulk quasiparticle spectra.

As a first result, we find that the $GW$ approximation provides a more consistent picture than KS-DFT: While the Rashba parameters differ substantially for LDA and GGA, they are brought back into good agreement by the quasiparticle correction, demonstrating a weak dependence on the reference KS system that is used as the starting point for the $GW$ calculation.

We show that the quasiparticle corrections have a significant effect on the band gap and the Rashba parameters which determine the spin splitting of the band-gap edge states. For BiTeI, the resulting $GW$ values of the bulk parameters are no longer in close agreement to the aforementioned ARPES measurements of Ref.~\onlinecite{NatMat_BiTeI}. We suggest an estimate in order to predict changes in the dispersion of the surface states induced by the many-body corrections. This estimate brings the theoretical and experimental data back into good agreement. In the case of BiTeCl, the $GW$ Rashba parameters are much closer to the experimental values than the KS results. There is only little difference between the $GW$ bulk parameters and the estimate for the surface state, and both fall in the error range of the experiment. The quasiparticle correction to BiTeBr yields a mixed success. While it improves on the bulk parameter $\alpha_R$, reflecting the ratio of the Rashba energy $E_R$ and the momentum offset $k_R$,\cite{Bahramy} with respect to LDA and GGA, the parameters $E_R$ and $k_R$ themselves are both underestimated, and the estimate for the surface state is no improvement in this case. Yet, the theoretical LDA$+GW$ and GGA$+GW$ values are again in very good agreement, while the corresponding LDA and GGA show notable differences.

\section{Computational details}

We employ the full-potential linearized augmented plane wave (FLAPW) method as implemented in the FLEUR code \cite{FLEUR} within both the
LDA of Ref.~\onlinecite{LDA_CA_PZ} and the GGA of Ref.~\onlinecite{GGA_PBE} for the exchange-correlation (XC) functional. We use these two approximations in order to gain insight into the effect of different reference one-particle band structures on results obtained within the one-shot $GW$ approximation realized by the SPEX code.\cite{SPEX} The DFT calculations were carried out with the use of a plane-wave cutoff of $k_{max}=4.0$ bohr$^{-1}$, an angular momentum cutoff of $l_{max}=12$, and a $7\times7\times7$ $\Gamma$-centered $k$-point sampling of the Brillouin zone (BZ). In order to treat quite shallow semi-core $d$-states, the FLAPW basis was extended by inclusion of conventional local orbitals.\cite{conv_LOs} To more accurately describe high-lying unoccupied states,\cite{Krasovskii_PRB_1997, Friedrich_PRBR_2011} one local orbital per angular momentum up to $l = 3$ was included for each atom. The linearization energy parameters for these local orbitals were set at the ``center of gravity'' of the respective partial density of states for each angular momentum $l$ within the energy interval from the Fermi level up to a higher energy of 90 eV. In order to document the quality of convergence, we report $GW$ results obtained with different numbers of states ($N_b$)  involved in the calculation of the dielectric matrix and the Green function and with two $\Gamma$-centered Monkhorst-Pack grids ($N_k\times N_k\times N_k$, where $N_k=5$ or 6). The dielectric matrix evaluated within the random-phase approximation (RPA) was represented with the use of the mixed product basis,\cite{Mix_product} where we chose an angular momentum cutoff in the muffin-tin spheres of 4 and a linear momentum cutoff of 3.5 bohr$^{-1}$. The spin-orbit interaction was included into the $GW$ calculations already at the level of the reference system. Not only the Green function but also the self-energy then acquire spin-off diagonal terms that lead to a many-body renormalization of the spin-orbit coupling, a phenomenon which is absent otherwise.

We consider all bismuth tellurohalides in the hexagonal crystal structure with the lattice parameters reported in Ref.~\onlinecite{Shevelkov_JSSC_1995} and the atomic positions obtained theoretically in Ref.~\onlinecite{BTX_PRL} by structural optimization. The crystal structure of the compounds is built up of alternating hexagonal layers Te-Bi-X stacked along the $c$ axis. Each three layers form a three-layer (TL) block, and the distance between these blocks is about one and a half times greater than interlayer distances within the block. BiTeBr is supposed to be in an ordered phase with the atomic order
similar to that in BiTeI but with Br instead of I (see also Ref.~\onlinecite{Eremeev_NJP_2013}). The structure of BiTeCl differs from that of iodide in the sense that the stacking order of Cl and Te layers alternate along the $c$ direction resulting in a doubling of the unit cell.\cite{Shevelkov_JSSC_1995}

\section{Results and discussion}

\begin{figure*}
\begin{center}
\includegraphics[width=\textwidth]{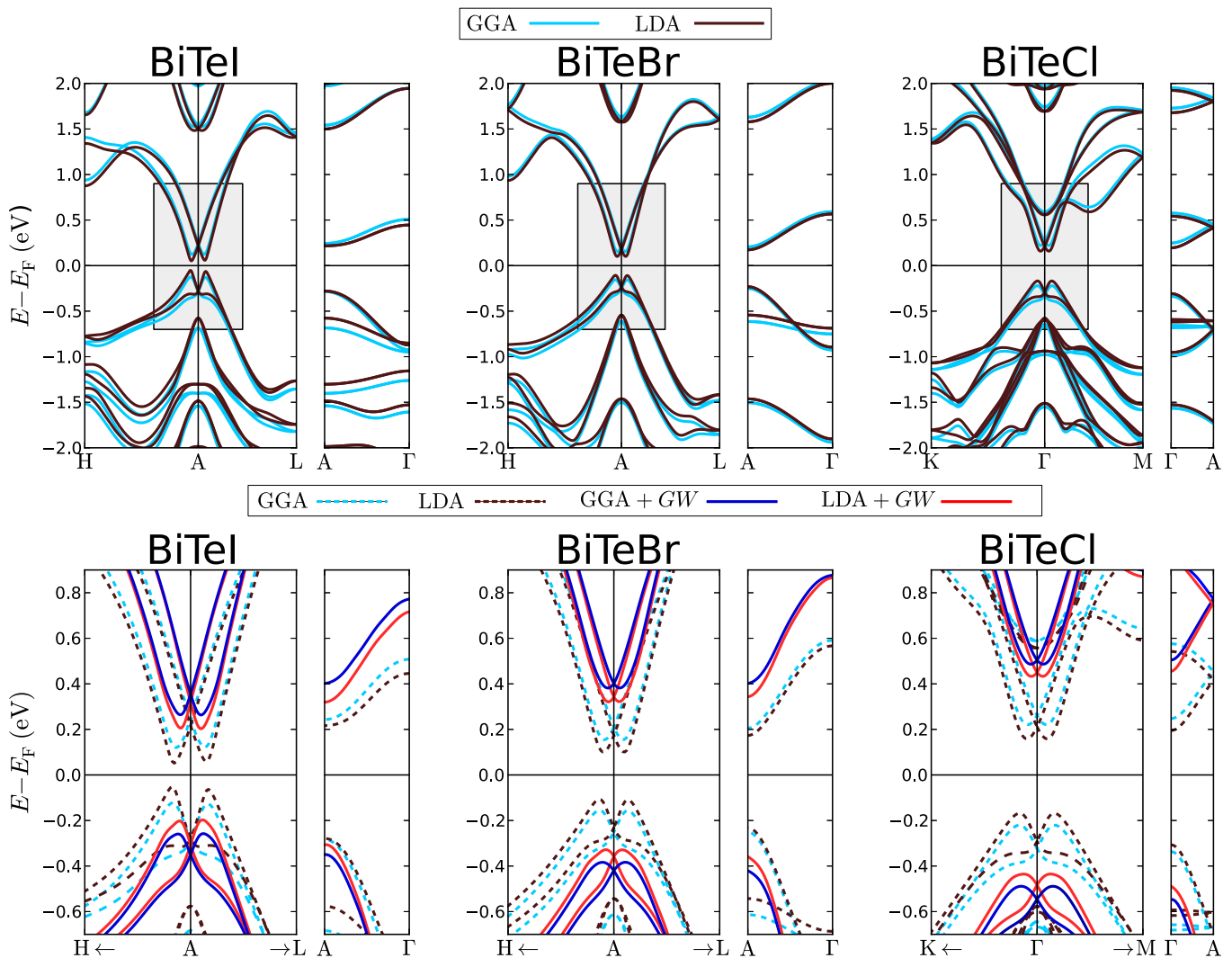}
\caption{(Color online) Upper row: Bulk band structure of bismuth tellurohalides, as obtained within Kohn-Sham DFT with the use of different approximations (LDA and GGA)  for the exchange-correlation functional. Shaded areas indicate the energy-momentum regions presented in the lower row. Lower row: The respective  quasiparticle spectra evaluated within the $GW$ approximation with the convergence parameters $N_k=5$ and $N_b=350$ (for BiTeCl the number of bands is  $N_b=580$). }
 \label{fig:ggalda}
\end{center}
\end{figure*}

The obtained LDA bulk band structures are presented in Fig.~\ref{fig:ggalda} (upper row). As seen in the figure, within the $H-A-L$ plane (or $K-\Gamma-M$ in BiTeCl, where the Brillouin-zone is folded along $\Gamma-A$ owing to the doubling of the unit cell) that is normal to the $\Gamma-A$ line the bands are
generally spin-split. Along the $\Gamma-A$ line, bands are spin degenerate and show a relatively weak dispersion compared to that along the mentioned plane. This reflects a quasi-two-dimensional character of the corresponding states, which is caused by the pronounced TL crystal structure of BiTeX. In the case of BiTeI and BiTeBr, the band gap is formed by the minima of the conduction band and the maxima of the valence band in the vicinity of the $A$ point of
the Brillouin zone (BZ). The parameters that characterize the Rashba-type spin splitting of the conduction band in the $A-L$ direction of the BZ in
BiTeI and BiTeBr are listed in Tables \ref{tab:bitei} and \ref{tab:bitebr}, respectively. These parameters, and also the band gap $E_{gap}$, vary
notably with the approximation chosen for the XC functional. The LDA gives systematically larger $k_R$, $E_R$, and $\alpha_R=2E_R/k_R$ and a smaller
$E_{gap}$ than those obtained with the use of the GGA (note that the inverse relation between $\alpha_R$ and $E_{gap}$ was revealed in Ref.~\onlinecite{Bahramy}). For BiTeI, the LDA and GGA values of $k_R$ are quite close to the experimental values found in Refs.~\onlinecite{NatMat_BiTeI} and \onlinecite{Sakano_Arxiv_2013}. Good agreement with experiment can also be found for the Rashba energy $E_R$ evaluated within the GGA. As to the parameter $\alpha_R$, which is quite sensitive to even small changes in $k_R$ and $E_R$, the GGA XC functional leads to a value that is in the error range of the experimental value $4.3\pm 0.9$ determined from the measured momentum offset and Rashba energy in Ref.~\onlinecite{Sakano_Arxiv_2013}. For BiTeBr, the DFT calculation gives Rashba parameters of which only $k_R$ is in the error range of the corresponding experimental value (see Table \ref{tab:bitebr}). In the case of BiTeCl (see Table \ref{tab:bitecl}), where the extrema of the valence and conduction bands appear at the $\Gamma$ point (see Fig. \ref{fig:ggalda}), the Rashba parameters are again larger within the LDA than those obtained within the GGA. The latter are somewhat closer to experiment, while, in fact, only the values $k_R$ and $\alpha_R$ fall within the error range of the experimental data.

\begin{table}
\caption{\label{tab:bitei} The Rashba parameters ($k_R$, $E_R$, $\alpha_R$) for the spin-orbit-split conduction band and the band-gap width ($E_{\mathrm{gap}}$) for BiTeI, which were obtained without (LDA and GGA) and with (LDA+$GW$ and GGA+$GW$) many-body corrections evaluated within the $GW$
approximation at different values of the convergence parameters $N_k$ and $N_b$. An estimate of the Rashba parameters for the surface state (see the text) is also presented. The experimental values taken from Ref.~\onlinecite{Sakano_Arxiv_2013} are shown with the respective error range represented by numbers in parentheses.}
\begin{ruledtabular}
\begin{tabular}{cccccc}
  $N_k/N_b$ & $E_{\mathrm{gap}}$,   & $E_R$, & $k_R$,       & $\alpha_R$,  \\
            & meV                    & meV     & 10$^{-3}$ \AA$^{-1}$   & \AA$\cdot$eV \\
\hline
\multicolumn{5}{c}{\textbf{GGA}}\\
           & 242            & 122 & 50   & 4.8 \\
\multicolumn{5}{c}{\textbf{GGA+$GW$}}\\
   5/350   & 520            & 93 & 37   & 5.1 \\
   5/500   & 512            & 93 & 37   & 5.1 \\
   6/350   & 501            & 93 & 37   & 5.1 \\
\multicolumn{5}{c}{\textbf{LDA}}\\
           & 104            & 159 & 56   & 5.6 \\
\multicolumn{5}{c}{\textbf{LDA+$GW$}}\\
   5/350   & 400            & 117 & 37   & 6.3 \\
\multicolumn{2}{l}{\textit{Estimate}} & 92     & 45     &  4.2    \\
\multicolumn{5}{c}{\textbf{Experiment}}\\
 & 380\footnotemark[1]    & 100\footnotemark[1]   & 52\footnotemark[1]    & 3.85\footnotemark[1] \\
 &             & 108(13)\footnotemark[2]   & 50(10)\footnotemark[2]    & 4.3(9)\footnotemark[2] \\
\end{tabular}
\footnotetext[1]{From Ref.~\onlinecite{NatMat_BiTeI}}
\footnotetext[2]{From Ref.~\onlinecite{Sakano_Arxiv_2013}}
\end{ruledtabular}
\end{table}

\begin{table}
\caption{\label{tab:bitebr} Same as in Table \ref{tab:bitei}, but for BiTeBr.}
\begin{ruledtabular}
\begin{tabular}{ccccc}
  $N_k/N_b$ & $E_{\mathrm{gap}}$,   & $E_R$, & $k_R$,       & $\alpha_R$,  \\
            & meV                    & meV     & 10$^{-3}$ \AA$^{-1}$   & \AA$\cdot$eV \\
\hline
\multicolumn{5}{c}{\textbf{GGA}}\\
                  & 310      & 55 & 34 & 3.3   \\
\multicolumn{5}{c}{\textbf{GGA+$GW$}}\\
            5/350 & 765      & 20 & 21 & 1.9   \\
            5/500 & 760      & 20 & 21 & 1.9   \\
            6/350 & 752      & 20 & 21 & 1.9   \\
\multicolumn{5}{c}{\textbf{LDA}}\\
                  & 207      & 73 & 37 & 4.0   \\
\multicolumn{5}{c}{\textbf{LDA+$GW$}}\\
            5/350 & 650      & 22 & 19 & 2.2   \\
\multicolumn{2}{l}{\textit{Estimate}} & 17     & 27     &  1.3   \\
\multicolumn{5}{c}{\textbf{Experiment}}\\
 &             & 42(10)\footnotemark[1]   & 43(10)\footnotemark[1]    & 2.0(7)\footnotemark[1] \\
\end{tabular}
\footnotetext[1]{From Ref.~\onlinecite{Sakano_Arxiv_2013}}
\end{ruledtabular}
\end{table}

\begin{table}
\caption{\label{tab:bitecl} Same as in Table \ref{tab:bitei}, but for BiTeCl.}
\begin{ruledtabular}
\begin{tabular}{ccccc}
  $N_k/N_b$ & $E_{\mathrm{gap}}$,   & $E_R$, & $k_R$,       & $\alpha_R$,  \\
         & meV                    & meV     & 10$^{-3}$ \AA$^{-1}$   & \AA$\cdot$eV \\
\hline
\multicolumn{5}{c}{\textbf{GGA}}\\
            & 441       & 41 & 33 & 2.5   \\
\multicolumn{5}{c}{\textbf{GGA+$GW$}}\\
   5/580    & 975       & 18 & 22 & 1.6   \\
   5/820    & 984       & 17 & 22 & 1.6   \\
   6/580    & 966       & 18 & 22 & 1.6   \\
\multicolumn{5}{c}{\textbf{LDA}}\\
            & 327       & 53 & 36 & 2.9   \\
\multicolumn{5}{c}{\textbf{LDA+$GW$}}\\
   5/580    & 868       & 18 & 20 & 1.8   \\
\multicolumn{2}{l}{\textit{Estimate}} & 15     & 26     &  1.2    \\
\multicolumn{5}{c}{\textbf{Experiment}}\\
 &             & 25(10)\footnotemark[1]   & 26(8)\footnotemark[1]    & 1.9(10)\footnotemark[1] \\
\end{tabular}
\footnotetext[1]{From Ref.~\onlinecite{Sakano_Arxiv_2013}}
\end{ruledtabular}
\end{table}

We performed calculations of the many-body corrections with different convergence parameters $N_k$ and $N_b$. In Table \ref{tab:diel_const} we present the corresponding RPA dielectric constants $\epsilon_{\infty}$, which gives an indication of how the screening properties depend on the convergence parameters. As is clearly seen in the table, with the use of the GGA band structure an increase of the number of bands at a fixed $N_k$ has practically no effect on
$\epsilon_{\infty}$. For BiTeBr, the dielectric constant does not show significant changes, when $N_k$ is increased at fixed number of bands. However, increasing $N_k$ at fixed $N_b$ has a stronger effect in BiTeI and BiTeCl: The dielectric constant grows by $\sim9\%$ for the former, while it decreases by
$\sim5\%$ for the latter. A similar situation is observed in the LDA case. An analysis has shown that for BiTeI the LDA-based RPA calculations of $\epsilon_{\infty}$ with $N_k=7$ and $N_b=350$ further increases the dielectric constant by $\sim5\%$. As we demonstrate below, the quasiparticle spectrum is much less sensitive to the changes in $N_k$ and $N_b$, and the fact that the dielectric constant is not fully converged with respect to the number
of \textbf{k}-points (except for the BiTeBr case) does not affect the conclusions made from the obtained results.

Now we discuss our results of the $GW$ calculations, which are based on the LDA or GGA electronic states as a reference one-particle band
structure. Fig. \ref{fig:ggalda} (lower row) represents the $GW$ results obtained with $N_k=5$ and $N_b=350$ (580 in the case of BiTeCl). It is worth noting that the changes in $N_k$ and $N_b$ have hardly an effect on the quasiparticle spectra, which is reflected in Tables \ref{tab:bitei}-\ref{tab:bitecl}. As
is clearly seen in the figure, the many-body corrections lead to a larger band gap, a smaller momentum offset, and a steeper dispersion along the $\Gamma-A$ line. Having a closer look at the values in the Tables \ref{tab:bitei}-\ref{tab:bitecl} the first observation is that the many-body quasiparticle corrections increase the band gap considerably with respect to the LDA and GGA values, bringing it very close to the experimental value indeed. In the case of BiTeI, the quasiparticle correction somewhat overestimates the experimental gap when the GGA reference is used, while the LDA+$GW$ value is nearly on top of the experimental one. Second, the GGA+$GW$ results listed in the tables demonstrate that the Rashba parameters are well converged. That is why we have performed the LDA+$GW$ calculations with $N_k=5$ and $N_b=350\,(580)$ only. Third, in the $GW$ calculations the difference between the Rashba parameters ($E_R$ and $k_R$) obtained with the use of different approximations to the XC functional is smaller than in the DFT calculations. Thus, we can infer with confidence that taking many-body corrections to the DFT band structure into account leads to a reduction of the Rashba energy and the momentum offset, which characterizes the Rashba-type spin-splitting of the bulk band-gap edges. The ratio of these parameters, as represented by $\alpha_R$, also demonstrates a reduction except for BiTeI, where it becomes larger.

In BiTeI, the obtained $GW$ results ``worsen'' the agreement with the experimental data. As a consequence, the interpretation done in Ref.~\onlinecite{NatMat_BiTeI} concerning the bulk nature of the Rashba-type spin-split state appears to be less legitimate in the light of the present $GW$ values. On the other hand, in BiTeCl the $GW$ corrections bring the theoretical and experimental values of the Rashba parameters into better agreement. The case of BiTeBr is a peculiar one. Here, the $GW$ values for the Rashba energy $E_R$ and momentum offset $k_R$ underestimate the experimental values by about the same factor so that their ratio, given by the $\alpha_R$ parameter, is in very close agreement with the experimental data.

\begin{table}
\caption{\label{tab:diel_const} The dielectric constant $\epsilon_{\infty}$ ($E\perp c$) obtained within the RPA for the considered semiconductors (for BiTeCl $N_b$ is shown in parentheses). The experimental value of $\epsilon_{\infty}$ for BiTeI is taken from Ref.~\onlinecite{BiTeI_diel_const}.}
\begin{ruledtabular}
\begin{tabular}{lccccc}
  XC       & $N_k$ & $N_b$ & BiTeI    &  BiTeBr   & BiTeCl  \\
\hline
GGA        & 5     & 350 (580)   & 16.2     & 14.7      & 15.1    \\
GGA        & 5     & 500 (820)   & 16.5     & 14.7      & 15.1    \\
GGA        & 6     & 350 (580)   & 17.6     & 14.8      & 14.4    \\
LDA        & 5     & 350 (580)   & 18.6     & 16.5      & 17.0    \\
LDA        & 6     & 350 (580)   & 19.9     & 16.5      & 16.2    \\
Exp.       &       &            & 19$\pm$2   &  &  \\
\end{tabular}
\end{ruledtabular}
\end{table}

Until the properties of quasiparticles in the surface state are calculated directly (which is a separate and very complicated problem at present), the presented results may serve as a basis for predicting the surface quasiparticle spectrum. Actually, as was experimentally established in Ref.~\onlinecite{Landolt} the offset $k_R$ in BiTeI found for the surface state is about $20\%$ larger than that for the bulk conduction band. The same change has been theoretically obtained in Ref.~\onlinecite{BiTeI_JETPLett}. Additionally, from DFT studies\cite{BTX_PRL, BiTeI_JETPLett} we know that the ratio between $\alpha_R$ for the surface state and that for the respective bulk state amounts to $0.66$. (Note that for BiTeCl one gets practically the same value, whereas for BiTeBr it is equal to 0.56.\cite{Eremeev_NJP_2013}) For consistency, a similar ratio for the Rashba energy $E_R$ should be equal to $0.79$. With these findings at hand, we can estimate Rashba parameters for the surface state at the BiTeI (0001) surface. Assuming the LDA+$GW$ values as a basis, we enlarge $k_R$ by $20\%$ and multiply $E_R$ and $\alpha_R$ by 0.79 and 0.66, respectively. As a result, we arrive at values listed in Table
\ref{tab:bitei} in the row marked as ``Estimate''. This estimate restores the good agreement, which was lost upon taking the $GW$ quasiparticle corrections into account. This can be interpreted in favor of a surface nature of the state experimentally observed in Refs.~\onlinecite{NatMat_BiTeI} and \onlinecite{Sakano_Arxiv_2013}.

Along with the dielectric constant and the band-gap width, the Rashba parameters estimated for the surface state in BiTeI indicate that in spite of the fact that at the DFT level, as a rule, the GGA gives more accurate results for the band structure than the LDA does, in pairs with the one-shot $GW$ approximation the latter becomes more preferable. Thus, with the use of the results reported in Refs.~\onlinecite{BTX_PRL}, \onlinecite{BiTeI_JETPLett}, and \onlinecite{Eremeev_NJP_2013} we can make a similar estimate for the surface states at the (0001) surface of BiTeBr and BiTeCl. As seen in Tables  \ref{tab:bitebr} and \ref{tab:bitecl}, the resulting $k_R$ and $E_R$ values do not cause a change of the situation: in BiTeCl the modifications occur within the error range of the experimental values, and in BiTeBr the discrepancy with the experiment remains. This discrepancy could be caused by the fact that we use the lattice parameters $a$ and $c$ reported in Ref.~\onlinecite{Shevelkov_JSSC_1995} for the disordered BiTeBr, where tellurium and bromine atoms randomly distributed within two layers facing Bi-atomic layer. An ordered BiTeBr has been grown very recently in Ref.~\onlinecite{Sakano_Arxiv_2013} by the chemical vapor transport method. However, in Ref.~\onlinecite{Sakano_Arxiv_2013} there is no information about the lattice parameters and atomic positions. We suppose that apart from surface effects a possible difference in the crystal structure can lead to the disagreement we observe.

\section{Conclusion}

In conclusion, we have presented quasiparticle spectra of bulk bismuth tellurohalides. We have shown that taking into account many-body corrections to the KS band structure from DFT calculations leads to a notable modification of the Rashba parameters ($E_R$, $k_R$, and $\alpha_R$), which characterize the spin-orbit splitting of the conduction-band minimum. We have revealed that the resulting $GW$ values of these parameters are well converged at a moderate \textbf{k}-point sampling of the BZ and number of bands and demonstrate a weak dependence on the approximation chosen for the exchange-correlation
functional. We have found that the $GW$ corrections improve the bulk band-gap value considerably with respect to the KS value. It should be noted that this
value turns out to be quite sensitive to the choice of the exchange-correlation functional on the level of KS DFT alone. We have shown that for BiTeI the $GW$ values of the Rashba parameters for the bulk conduction band worsen the good agreement with experiment that was reached at the DFT level. Thus, in the light of the $GW$ results, the interpretation of the state observed by the ARPES in Ref.~\onlinecite{NatMat_BiTeI} to be of a bulk nature appears less likely. The suggested estimate of these parameters for the surface state has restored the good agreement, which may indicate a surface nature of the state in question. We have also found that in the case of BiTeCl the experimental data cannot be reproduced by DFT. However, both the $GW$ values of the bulk Rashba parameters and estimates for those of the surface state match the experimental values within the error bars of the experiment. As to BiTeBr, neither the DFT nor the $GW$ approximation and the further estimate is able to reproduce simultaneously the available experimental data on $E_R$, $k_R$, and $\alpha_R$.

\section*{\label{sec:acknowledgments}Acknowledgments}

We acknowledge partial support from the Basque Country Government, Departamento de Educaci\'{o}n, Universidades e Investigaci\'{o}n (Grant No. IT-366-07), and the Spanish Ministerio de Ciencia e Innovaci\'{o}n (Grant No. FIS2010-19609-C02-00).

\end{document}